\documentstyle[epsf]{lamuphys}
\makeatletter
\let\chapter\hid@chapter
\makeatother
\begin{document}
\pagenumbering{arabic}
\title{The Hubble Deep Field-South QSO}

\author{Katrina\,Sealey\inst{1}, Michael\,Drinkwater\inst{1},
John\,Webb\inst{1}, Brian\,Boyle\inst{2}}

\institute{Physics, University of New South Wales, Sydney 2052, Australia
\and
Anglo-Australian Observatory, PO Box 296, Epping 2121, Australia}

%\authorrunning{M.\,Drinkwater et al.}
\maketitle

\begin{abstract}
The Hubble Deep Field-South was chosen to have a QSO (RA 22:33:37.6
Dec $-$60:33:29 J2000 and B=17.5) in the field to allow for studies of
absorption systems intersecting the sight line to the QSO.  To assist
in the planning of HDF-S observations we present here a ground-based
spectrum of the QSO. We measure a redshift of $z=2.24$ for the
quasar and find associated absorption in the spectrum at $z=2.204$ as
well as additional absorption features.
\end{abstract}

\section{Introduction}

Unlike the original Hubble Deep Field, the
Hubble Deep Field South (HDF-S) was chosen specifically to contain
a $z>2$ QSO suitable for studying the relationship
between the high redshift galaxies identified in the HDF-S and the
absorption lines in the spectrum of the HDF-S QSO. The QSO was found
on a UK Schmidt Telescope objective prism plate scanned by Mike Irwin
using the Automated Plate Measuring facility in Cambridge analysed by
Paul Hewett and then confirmed by observations at the Anglo-Australian
Telescope (Boyle 1997). To aid future observations of the HDF-S we
present here a low-resolution spectrum of the QSO; the data
are available at http://bat.phys.unsw.edu.au/$\sim$kms/hdfs/.

\section{The QSO Spectrum}

We observed the HDF-S QSO on 1997 October 21 with the Australian
National University 2.3m Telescope at Siding Spring with $\approx$~2
arcsec seeing. The Double Beam Spectrograph was used with a 300~l/mm
grating in the blue arm (binned by 2 in dispersion), and a 158~l/mm
grating in the red giving a resolution of 9~\AA~in each arm. A
dichroic was used to split the light at $\approx$~5400--5500~\AA\
(there is some uncertainty in the spectrum in this region). The
spectrum shown in Fig.~1 is the weighted average of 2 exposures (1200s
and 900s).

\subsection{Emission Lines and QSO Continuum}
The spectrum has emission lines due to Lyman-$\alpha$ + NV, SiIV +
OIV] , CIV, CIII] and MgII.  Broad FeII emission is seen around
7200--8700\AA.  The redshift of the QSO derived by simultaneously
fitting to CIV, CIII] and MgII is $z=2.24$.  The individual redshifts
for each line are z=2.23 (CIII] and CIV) and z=2.25 (MgII); the
redshift difference is not unusual (Espey et al.\ 1989). A power law
fit to the QSO continuum, avoiding obvious emission features (both
broad and narrow) gives a slope of $\alpha=-0.8\pm0.1$ (where
$f_\nu\propto\nu^\alpha$) and is shown in Fig.~1.
%\raisebox{-2cm}[0cm][0cm]{\makebox[0cm][c]
%{\hspace{-7cm}\parbox{12cm}{\em Paper presented at the ESO/Australia Workshop:
%Looking Deep in the Southern Sky, held in Sydney, 1997 December 10-12,
%eds. R. Morganti et al.  }}}

\subsection{Absorption Systems}

An initial investigation of the QSO shows absorption present from
intervening systems.
There appears to be ``associated'' absorption at $z_{abs} = 2.204$,
suggested by the strong, narrow absorption lines in the blue wing of
the CIV and NV emission lines (the latter falls directly between the
Lyman-$\alpha$ and NV emission peaks).  The blended Lyman-$\alpha$ + NV
emission profile drops very sharply from Ly$\alpha$ towards NV
(compared to the blue wing of Lyman-$\alpha$, which must be cut into by
multiple Lyman-$\alpha$ systems), probably due to absorption, rather than
the 2 emission lines being resolved.  The absorption line is most
likely to be NV because of the presence of strong CIV absorption at
the same redshift.  If the above interpretation is correct, there are
at least 3 additional strong absorption lines (the 2 conspicuous lines
straddling the SiIV emission line and one mid-way between the NV 
and SiIV emission lines) which we can not yet identify
reliably. If the lines are interpreted to be CIV we are seeing a 
rich complex of CIV absorption. These lines nevertheless
reveal additional structure in an extensive system of absorbers,
which may arise in a cluster of galaxies around the QSO.  

\begin{figure}
%\vspace{2.5cm}
\centering
\epsfxsize=12.8cm\epsfbox[42 160 592 420]{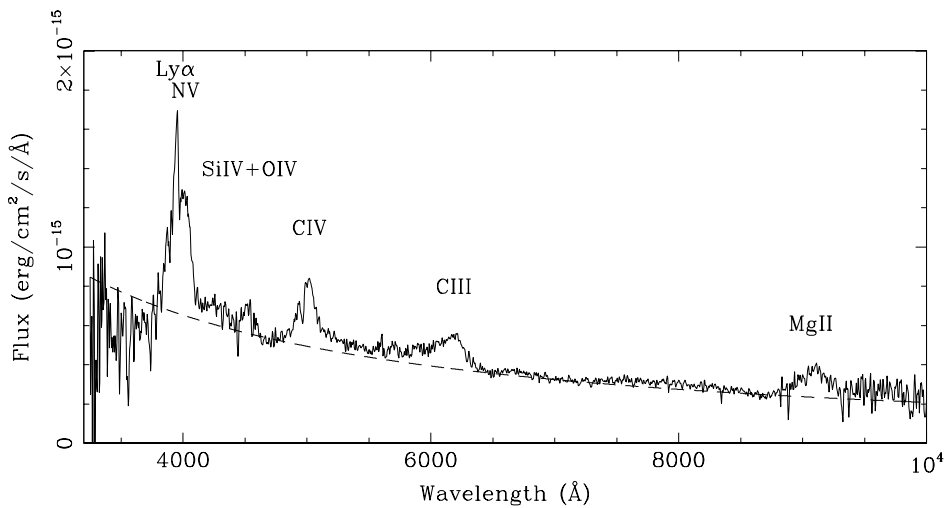}
\caption{Spectrum of the HDF-South QSO.}
\end{figure}
 
% ---- Bibliography ----
%

Note: Katrina Sealey's new address is Macquarie University, NSW 2109, 
Australia.
\end{document}